\documentclass[12pt]{article}
\pdfoutput=1 

\usepackage{jheppub} 

\usepackage[T1]{fontenc} 
\usepackage{tikz} 
\usetikzlibrary{calc,through,backgrounds,patterns,decorations.pathmorphing,decorations.markings,%
		trees,positioning,arrows,chains,shapes.geometric,%
    decorations.pathreplacing,shapes}

\usepackage{pgfplots}
\usepgfplotslibrary{fillbetween}    
\pgfplotsset{compat=1.10}

\usepackage{amsmath}
\usepackage{multirow}
\usepackage{array}

\usepackage{enumerate}

\usepackage[export]{adjustbox}

\relax
\def\be{\begin{equation}}
\def\ee{\end{equation}}
\def\bea{\begin{eqnarray}}
\def\eea{\end{eqnarray}}

\newcommand{\bsea}{\begin{subeqnarray}}
\newcommand{\esea}{\end{subeqnarray}}
\newbox\pippobox

\def\6{\partial}

\def\a{\alpha}

\def\sq
\def\a{\alpha}

\numberwithin{equation}{section}

\title{\boldmath Fixing the non-relativistic expansion of the 1PM potential} 

\author[a ]{Gianluca Grignani,}
\author[b]{Troels Harmark,}
\author[a,b]{Marta Orselli}
\author[a,b ]{Andrea Placidi,} 

\affiliation[a]{Dipartimento di Fisica e Geologia, Universit\`a di Perugia,
INFN Sezione di Perugia, 
Via A. Pascoli, 06123 Perugia, Italia}
\affiliation[b]{ Niels Bohr Institute, Copenhagen University,  Blegdamsvej 17, DK-2100 Copenhagen \O{}, Denmark
}

\emailAdd{ gianluca.grignani@unipg.it}
\emailAdd{harmark@nbi.ku.dk}
\emailAdd{orselli@nbi.dk }
\emailAdd{andrea.placidi@studenti.unipg.it}

\date{}

\preprint{}

\abstract{
   
We obtain a first order post-Minkowskian two-body effective potential whose post-Newtonian expansion directly reproduces the Einstein-Infeld-Hoffmann potential. Post-Minkowskian potentials can be extracted from on-shell scattering amplitudes in a quantum field theory of scalar matter coupled to gravity. Previously, such potentials did not reproduce the Einstein-Infeld-Hoffmann potential without employing a suitable canonical transformation. In this work, we resolve this issue by obtaining a new expression for the first-order post-Minkowskian potential. This is accomplished by exploiting the reference frame dependence that arises in the scattering amplitude computation. Finally, as a check on our result, we demonstrate that our new potential gives the correct scattering angle.
}


\begin{document}

\maketitle
\flushbottom




\section{Introduction}

The calculation of two-body effective potentials for the conservative dynamics of binary systems from on-shell scattering amplitudes is, by now, a well established technique. After its debut~\cite{Iwasaki:1971vb} in the Post-Newtonian (PN) perturbative framework of General Relativity~\cite{Blanchet:2013haa}, this method has been mostly employed in the context of the Post-Minkowskian (PM) expansion~\cite{Bjerrum-Bohr:2018xdl,Cheung:2018wkq,Cristofoli:2019neg,Cristofoli:2020uzm}. Indeed, since this technique does not require a non-relativistic (NR) expansion on the typical velocity of the system $v$, a Post-Minkowskian approach where the Newton's constant $G$ is identified as the one and only perturbative parameter of the theory, allows us to obtain the most comprehensive two-body interaction potential that we could build from scattering amplitudes. 

In dealing with gravitationally bound states like compact binaries, the virial theorem tells us that kinetic and potential energies must share the same order of magnitude, i.e.\
\begin{equation*}
	\frac{GM}{r c^2} \sim \frac{v^2}{c^2}~,
\end{equation*}
$M$ being the total mass of the system.
Therefore, there is no straightforward way to determine whether keeping all order in velocity would effectively bring an increase in accuracy. Nevertheless, PM results for the conservative two-body dynamics have proved to be quite promising~\cite{Antonelli:2019ytb}, both as an input for effective one-body models and as a mean to crosscheck and supplement the preexisting PN results. Specifically in view of the latter purpose, it is important to elucidate how the PM potentials behave once non-relativistically expanded, especially if we want to extract out of them the corresponding terms in the PN expansion. With the intention of clarifying this aspect of PM calculations from scattering amplitudes in the simplest possible conditions, this work will be focused on the NR limit of the first order PM potential. 
 
We will start by briefly reviewing how PM potentials are obtained in general, pointing out the mismatch between the NR limit of present 1PM potentials and the associated PN results. Then we will show how the reference frame dependence of the potential can be exploited to find a new expression for the 1PM potential that, once expanded for small velocities, directly reproduces the well-known Einstein-Infeld-Hoffmann 1PN potential~\cite{10.2307/1968714}. Extending this nice feature to higher PN orders seems a more challenging task and we leave it for future work. Here we will instead reproduce higher PN orders by means of a simple canonical transformation.
Finally we will check the physical consistency of our new 1PM potential against the ones already present in the literature, by showing that from it one can obtain the correct scattering angle (see for example \cite{Bern:2019crd,Cristofoli:2019neg,PhysRevD.100.066028}). 

The computation of the scattering angle for non-relativistic quantum mechanical Hamiltonians has a long history. Typically, the interest has been mainly focused on finding approximate (semi-classical) solutions, first through the WKB-approximation, later by considering the eikonal limit (see, e.g., refs. \cite{Sugar:1969rn,Paliov:1974fu,Wallace:1973iu}). Amplitudes methods have been used more recently to compute scattering angles~\cite{KoemansCollado:2019ggb} and we will adopt here the approach of ref.~\cite{Bjerrum-Bohr:2019kec}.

In this paper, following the vast majority of works on this subject, we leave aside all the finite-size~\footnote{See ref.~\cite{Goldberger:2004jt} and the recent ref.~\cite{Cheung:2020sdj} for works on how to include finite-size effects.} and spin effects~\footnote{See Refs.~\cite{Chung:2018kqs,Guevara:2018wpp} for recent attempts to include the spin.}, that could fit the proper description of some astrophysical binary systems.

\section{Post-Minkowskian potential}

Within the PM scheme, the general two-body effective Hamiltonian for a system of two compact objects gravitationally bound to each other, in the point-particle approximation  can be written as
\begin{equation}
	H_{\text{PM}}(\boldsymbol{p}_1, \boldsymbol{p}_2, \boldsymbol{r}) = \sqrt{\boldsymbol{p}_1 c^2 + m_1^2 c ^4} + \sqrt{\boldsymbol{p}_2 c^2 + m_2^2 c ^4} + \sum_{n=1}^{+\infty} V_{n\text{PM}}(\boldsymbol{p}_1, \boldsymbol{p}_2, \boldsymbol{r})~,
\end{equation}
where an effective gravitational potential organized in a PM expansion (i.e.\ in powers of $G$) is added to the free-particle energies. Momenta and masses of the two objects are labeled as $\boldsymbol{p}_{1,2}$ and $m_{1,2}$ while $\boldsymbol{r} \equiv \boldsymbol{x}_1 - \boldsymbol{x}_2$ stands for their radial separation. At the 1PM order and selecting the center of mass reference frame we get 
\begin{equation}
	H_{1\text{PM}}(\boldsymbol{p}, \boldsymbol{r}) = \sqrt{\boldsymbol{p} \, c^2 + m_1^2 c ^4} + \sqrt{\boldsymbol{p} \, c^2 + m_2^2 c ^4} + V_{1\text{PM}}(\boldsymbol{p}, \boldsymbol{r}),
\end{equation}
with $\boldsymbol{p} \equiv \boldsymbol{p}_1 = -\boldsymbol{p}_2$.
The interaction potential represents the only non-trivial part of the Hamiltonian. Its determination at any given PM order can be accomplished within the framework of on-shell scattering amplitudes~\cite{Bjerrum-Bohr:2018xdl,Cheung:2018wkq,Cristofoli:2019neg,Cristofoli:2020uzm} and we will now describe the main features of this derivation.

The underlying theory, upon which the whole method is based, is a quantum field theory of gravity where in general gravitons are coupled to quantum fields whose degrees of freedom reflect the properties of the systems we wish to describe. In particular, the standard choice for the action is~\cite{Bjerrum-Bohr:2018xdl}
\begin{equation} \label{Action}
	\mathcal{S}=\int d^4 x \, \sqrt{-g} \bigg[ \frac{R}{16 \pi G} + \frac{1}{2} \sum_{a=1,2}\big( g^{\mu \nu} \partial_{\mu} \phi_a \partial_\nu \phi_a - m^2_a \phi^2_a \big) \bigg].
\end{equation}
Here the pure gravity sector can be limited to the Einstein-Hilbert term without renormalizability concerns, since the classical, low energy physics we are interested in is unaffected by higher order terms in the curvature. The remaining parts introduce a minimal gravity-matter coupling in terms of two real scalar fields $\phi_{1,2}$, with masses $m_{1,2}$ corresponding to the two astrophysical objects we want to describe. The minimal coupling relies on the point-particle approximation and is effectively meaningful only when dealing with really compact objects like black holes~\footnote{Even for relatively compact objects such as neutron stars, finite-size effects~\cite{Goldberger:2004jt,Cheung:2020sdj} start to be important~\cite{Hinderer:2007mb}.}. Moreover, the fact that we work with scalar fields specializes our description to compact binaries whose black holes are Schwarzschild black holes~\footnote{The extension of this approach to Kerr black holes requires the inclusion of spin effects~\cite{Chung:2018kqs,Guevara:2018wpp}.}.


As for the metric, the standard quantization scheme involves a perturbative expansion around the flat Minkowskian metric $\eta_{\mu \nu}$, namely
\begin{equation} \label{LinerizedMetric}
\begin{split}
	g_{\mu \nu} & = \eta_{\mu \nu}+\sqrt{32 \pi G} h_{\mu \nu}~,\\
    g^{\mu \nu } & = \eta^{\mu \nu} -\sqrt{32 \pi G} h^{\mu \nu} + 32 \pi G h^{\mu \rho} h_{\rho}^{\phantom{\rho} \nu} + O(\sqrt{G^3})~,
\end{split}
\end{equation}
where the small fluctuation $h_{\mu \nu}$ is identified as the graviton field. 

After a gauge fixing choice, which is typically the De Donder gauge corresponding to adding the following additional term in the action
\begin{equation}
	S_{\text{GF}}=  \int d^4 x \, \bigg( \partial^{\rho} h_{\rho \alpha} - \frac{1}{2}\partial_{\alpha} h \bigg)^2~,
\end{equation}
the Feynman rules of this theory can be unambiguously derived (they can be found for example in ref.~\cite{Holstein:2008sx}), thus allowing for the diagrammatic computation of the scattering amplitudes. 

The relevant process for the derivation of the potential is a scattering process involving only matter fields, {\it i. e.} of the type
\begin{equation} \label{RelevantProcess}
	\phi_1 \phi_2 \rightarrow \phi_1 \phi_2,
\end{equation}
with one or more gravitons in the internal propagators. Not all the terms in the amplitudes relative to this process are needed, but only those that encode a long-range and classical interaction. More precisely, the long-range condition is satisfied only when a pole in $q^2$ is present, where $q$ is the momentum transfer. Analytical terms in $\bf{q}$ would provide local or ultra-local contributions to the potential, namely proportional to a Dirac delta or its derivatives. The classical requirement is somewhat more subtle and entails selecting those components that depend on the dimensionless ratios $m_{1,2}/\sqrt{q}$. This is due to a compelling cancellation of $\hbar$ that arises in the loop expansion of processes involving an interplay between massive and massless particles, whenever the aforementioned ratio is present. For more details on this $\hbar$ cancellation we refer to the thorough analysis of ref.~\cite{Kosower:2018adc}.

These conditions greatly simplify the task of computing the required classical contributions of the quantum amplitude. Moreover, they enable the use of modern techniques for the evaluation of amplitudes in the spinor-helicity formalism, whose foundations are generalized unitarity~\cite{Bern:1994zx} and double-copy relations \cite{Bern:2010ue}. In this scheme, gravity tree amplitudes are obtained from the simpler gauge-theory ones and then they are employed to evaluate the singular part of the corresponding loop amplitudes. In ref.~\cite{Bern:2019crd} such methods have been used to compute gravity amplitudes up to two loops.

The specific link between those amplitudes and the potential can be established through the following equivalent procedures:
\begin{itemize}
	\item Matching order by order the on-shell amplitudes of the full theory discussed above with those that come from an effective field theory of two NR scalar fields $A$ and $B$ described by the Lagrangian
	\begin{equation}
		\begin{split}
		\mathcal{L} & = \int \frac{d^3 \boldsymbol{k}}{(2 \pi)^3} \, \bigg[A^{\dagger}(-\boldsymbol{k})\bigg(i \partial_{t} -\sqrt{\boldsymbol{k}^2+m^2_{A}} \bigg) A(\boldsymbol{k}) + B^{\dagger}(-\boldsymbol{k})\bigg(i \partial_{t} -\sqrt{\boldsymbol{k}^2+m^2_{B}} \bigg) B(\boldsymbol{k})\bigg] + \\
		& + \int \frac{d^3 \boldsymbol{k}}{(2 \pi)^3} \frac{d^3 \boldsymbol{k'}}{(2 \pi)^3} \, V(\boldsymbol{\boldsymbol{k}, \boldsymbol{k}'}) A^{\dagger}(\boldsymbol{k})A(\boldsymbol{k}') B^{\dagger}(-\boldsymbol{k}') B(-\boldsymbol{k})~,
		\end{split}
	\end{equation}
	where $V(\boldsymbol{\boldsymbol{k}, \boldsymbol{k}'})$ is the sought potential.
	This was originally proposed in ref.\ \cite{Cheung:2018wkq}, in the first PM potential derivation from scattering amplitudes, and then in ref.\ \cite{Bern:2019crd} it has been employed to compute up to the 3PM potential.
	
	\item Adapting the Born series formalism, that is commonly used in NR quantum mechanics, to the relativistic case. Under the conditions at hand for the considered scattering process, this generalization can be easily accomplished. If we label the initial and final scattering states as $| p_1, p_3 \rangle$ and $| p_2, p_4 \rangle$ respectively, the Born series that gives the momentum space potential reads
	\begin{equation} \label{BornS}
	\begin{split} 
	\langle p_2, p_4| V | p_1, p_3 \rangle & = M(p_1, p_2,p_3, p_4) + \\ & - 	\lim_{\varepsilon\to 0^+} \int \frac{d^3 \boldsymbol{k}_1}{(2\pi)^3}\frac{d^3 \boldsymbol{k}_2}{(2\pi)^3} \, \frac{M(p_1, p_3,k_1, k_2) 	M(k_1, k_2,p_2, p_4)}{E_{\boldsymbol{p}_1} + E_{\boldsymbol{p}_3}- E_{\boldsymbol{k}_1} - E_{\boldsymbol{k}_2} + i\varepsilon} + ... ~,
	\end{split}
	\end{equation}
	where $M$ is the full on-shell amplitude of the process (\ref{RelevantProcess}) and $k_i= \big(E_{\boldsymbol{k}_i} / c, \boldsymbol{k}_i\big)$. Then the potential at any order $N$ in the PM expansion can be determined by extracting the $O(G^N)$ contributions in the right hand side of eq.\ (\ref{BornS}), according to the simple correspondence 
	\begin{equation*}
	N\text{-loop amplitude} \iff O(G^{N+1}) \text{ contributions} .
	\end{equation*}
	This method was first introduced in ref.~\cite{Cristofoli:2019neg} and used to compute the potential up to the 2PM order.
\end{itemize}
Indeed in both these schemes the amplitudes are needed only in their long-range and classical parts.

Finally, the resulting potential in momentum space can be recasted in its usual position space expression with a Fourier integration on $\boldsymbol{q}$, the spatial component of the momentum transfer.

\subsection{The 1PM Potential}

Let us now focus on the 1PM potential. In this approximation the link amplitudes-potential is quite simple: the long-range, classical part of the tree-level amplitude corresponds directly to the momentum space expression of the potential. From a diagrammatic point of view, since there are no annihilation channels to deal with, this tree amplitude consist of a single Feynman diagram that one can readily evaluate:
\begin{equation} \label{TreeAmp}
	M^{\text{tree}}(p_1,p_2,p_3,p_4) =\includegraphics[valign=c,width=0.30\linewidth]{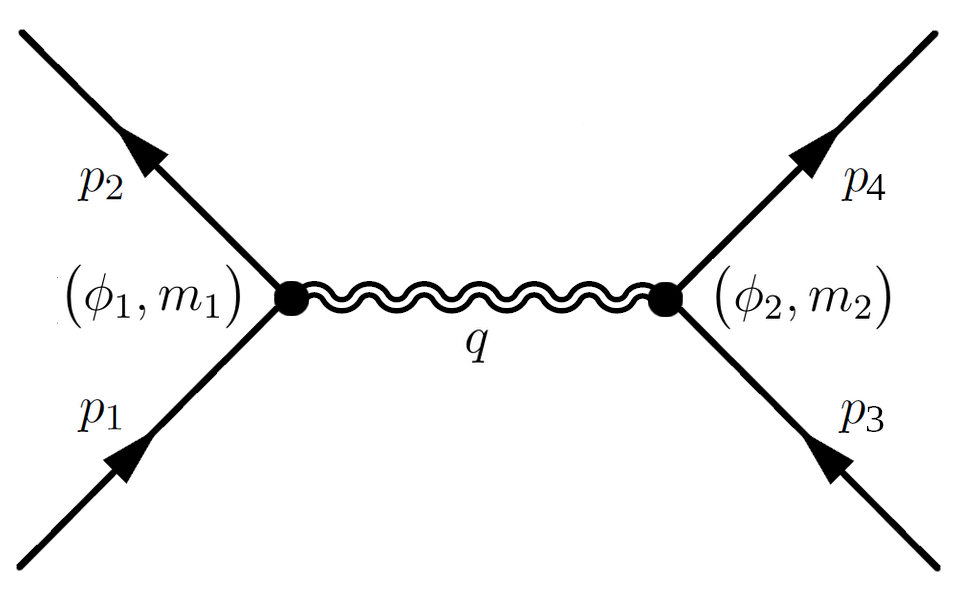}= \frac{4 \pi G}{\sqrt{E(p_1)E(p_2)E(p_3)E(p_4)}} \frac{C(p_1, p_2, p_3, p_4)}{q^2}~,
\end{equation}
where the contraction $C$ between the pair of scalar-graviton vertices and the numerator of the graviton propagator turns out to be
\begin{equation}
	\begin{split}
	C(p_1, p_2, p_3, p_4) & = (p_1 \cdot p_3)(p_2 \cdot p_4) + (p_1 \cdot p_4)(p_2 \cdot p_3) - (p_1 \cdot p_2)(p_3 \cdot p_4) + \\
	& + (p_3 \cdot p_4) m^2_1 c^2 +
	(p_1 \cdot p_2) m^2_2 c^2 - 2m^2_1 m^2_2 c^4~.
	\end{split}
\end{equation}

The next step is the Fourier transform that leads to the position space potential. While the on-shell amplitude is clearly reference frame invariant, the Fourier transform is not, since in performing it, it becomes important how 4-momenta and energies actually depend on $\boldsymbol{q}$.

Usually  Fourier transforms are carried out in the center of mass reference frame with the momentum transfer defined through:
\begin{equation}\label{4mom}
	q^\mu \equiv (p_1)^\mu - (p_2)^\mu = (p_4)^\mu - (p_3)^\mu = (0, \boldsymbol{q})~,
\end{equation}
with a zero time component compatible with energy conservation. With this choice the denominator of the graviton propagator becomes simply
\begin{equation}\label{denprop}
	\frac{1}{q^2} = - \frac{1}{|\boldsymbol{q}|^2}~.
\end{equation}
At the same time, external 4-momenta are written as
\begin{equation}
	\begin{split}
	(p_{1,3})^\mu & = (E_{1,2}, \pm \boldsymbol{p})~,  \qquad \text{[incoming]}\\
	(p_{2,4})^\mu & = (E_{1,2}, \pm \boldsymbol{p}')~, \, \, \, \, \, \, \, \, \, \, \, \text{[outgoing]}
	\end{split} 
\end{equation}
where
\begin{equation}
	E_{1,2} = c\sqrt{\boldsymbol{p}^2 + m_{1,2}^2 c^2}~, \quad \boldsymbol{p}'=\boldsymbol{p}-\boldsymbol{q}~.
\end{equation}
In other words the dependence of the external 3-momenta on $\boldsymbol{q}$, which then reflects itself on 4-momenta and energies, is limited to the final states:
\begin{equation} 
\begin{cases} \label{Standardframe}
\boldsymbol{p}_1 = - \boldsymbol{p}_3 = \boldsymbol{p} \\ 
\boldsymbol{p}_2 = - \boldsymbol{p}_4 = \boldsymbol{p} - \boldsymbol{q}
\end{cases}.
\end{equation}
As we will see, in the center of mass reference frame this is not the only possible choice of momenta.
Under these conditions one has
\begin{equation}
\bigg[M^{\text{tree}}(\boldsymbol{q})\bigg]_{\substack{
		\text{long-range} \\  \text{classical} \phantom{ab} }} = \dfrac{4 \pi G}{E_1 E_3} \dfrac{m^2_1 m^2_2 c^4 - 2 (p_1 {\cdot} p_3)^2}{|\boldsymbol{q}|^2}.
\end{equation}
The only $\boldsymbol{q}$-dependent quantity left is the absolute square in the denominator, so that the simple Fourier transform
\begin{equation}
\int \frac{d^3 \boldsymbol{q}}{(2 \pi )^3} \, \frac{e^{-i \boldsymbol{q} \cdot \boldsymbol{r}}}{|\boldsymbol{q}|^2} = \frac{1}{4 \pi r}
\end{equation}
gives directly the position space 1PM potential
\begin{equation} \label{Vlett}
	V^{[\text{lit.}]}_{1\text{PM}}(\boldsymbol{p}, \boldsymbol{r}) = - \frac{G }{E_1 E_2} \frac{2(p_1 \cdot p_3)^2 - m_1^2 m_2^2 c^4 }{r},
\end{equation}
where, with respect to the binary system of astrophysical bodies, we can regard $\boldsymbol{p}$ as their momentum in the center of mass frame and $r$ as their radial separation. This is the 1PM result for the potential on which refs.~\cite{Cheung:2018wkq,Bern:2019crd,Cristofoli:2019neg} agree upon.

\subsection{Issues With the NR Limit of the PM potential}

The 1PM potential (\ref{Vlett}) contains all orders of $1/c^2$ in the $c\rightarrow \infty$ expansion while being at the first order in the Newton's constant $G$. Therefore, one would expect its NR expansion should reproduce all the $O(G)$ terms of the potential encompassed by the PN Hamiltonian currently used in gravitational wave astronomy~\cite{Jaranowski:2015lha,Damour:2014jta}. At present the PN Hamiltonian has been computed up to 5.5PN order~\cite{Bini:2020wpo}.

However, if we take the result (\ref{Vlett}) and evaluate its NR expansion up to order $O(1/c^2)$ we find 
\begin{equation}\label{Vlettexp}
	V^{[\text{lit.}]}_{1\text{PM}}(\boldsymbol{p}, \boldsymbol{r}) = -\frac{G m_1 m_2}{r} - \frac{G m_1 m_2}{2 r c^2}\bigg(3 \frac{\boldsymbol{p}^2}{m_1^2} + 3 \frac{\boldsymbol{p}^2}{m_2^2} + 8 \frac{\boldsymbol{p}^2}{m_1 m_2} \bigg) + O(1/c^4)~.
\end{equation}
We can then compare this result to the well-know 1PN potential, first found by Einstein-Infeld-Hoffmann in ref.\ \cite{10.2307/1968714}, namely
\begin{equation} \label{EIHpot}
	V_{1\text{PN}} = - \frac{G m_1 m_2}{r} - \frac{G m_1 m_2}{2rc^2} \bigg[\frac{3 \boldsymbol{p}^2}{m^2_1} + \frac{3 \boldsymbol{p}^2}{m^2_2} + \frac{7\boldsymbol{p}^2}{ m_1 m_2} + \frac{(\boldsymbol{p} \cdot \hat{\boldsymbol{r}})^2}{m_1 m_2} \bigg] + O(G^2)~.
\end{equation}
As it clearly appears, these two results are generally different, unless one assumes $\boldsymbol{p}$ to be directed along $\hat{\boldsymbol{r}}=\frac{\boldsymbol{r}}{r}$, the radial separation direction. This apparent mismatch does not imply that the potential \eqref{Vlettexp} is incorrect. Indeed, it is well-known that the two potentials  \eqref{Vlettexp} and \eqref{EIHpot}  are related by a canonical coordinate remapping. More generally, it is discussed in ref.\ \cite{Bern:2019crd} how a suitable canonical transformation can establish the consistency between their 3PM potential and the 4PN one.

Nevertheless, in virtue of the theoretical possibility of finding unprecedented PN results from the NR expansion of the PM ones, it would be preferable to have PM potentials which present {\it a priori} a NR behaviour consistent with what has been already computed in the PN framework. In the following section we will show how the the 1PM potential can be modified in this sense, while leaving untouched the De Donder gauge fixing.

\section{A New Expression For the 1PM Potential}

We start  with a modified expression of the four momentum transfer, still compatible with the conservation laws. Instead of \eqref{4mom} we use~\footnote{This is a particular case of the generalized propagator introduced by Hiida and Okamura in ref.~\cite{Hiida:1972xs}, and later taken into account also in ref.~\cite{BjerrumBohr:2002kt}. According to their notation, we are considering the propagator which follows after having fixed the gauge parameter x to 1.}
\begin{equation}
	q = \bigg(\dfrac{\sqrt{(E_1 - E_2)(E_4 - E_3)}}{c}, \boldsymbol{q} \bigg)~.
\end{equation}
Here the time component is written in a form that satisfies in general energy conservation and that is symmetric with respect to the pair of particles involved in the scattering process. The denominator of the graviton propagator \eqref{denprop} becomes
\begin{equation}
\frac{1}{q^2} = \frac{1}{q_0^2 - |\boldsymbol{q}|^2} = \dfrac{c^2}{\big[E(p_1) - E(p_2)\big]\big[ E(p_4) - E(p_3) \big] - |\boldsymbol{q}|^2 c^2}~.
\end{equation}
Surprisingly, this is enough to significantly change the expression of the long-range classical part of the tree level amplitude. In fact, if we specify the 3-momenta as in (\ref{Standardframe}), we end up with
\begin{equation} \label{StandardFrameAmp}
	\begin{split}
	\bigg[M^{\text{tree}}(\boldsymbol{q})\bigg]_{\substack{
			\text{long-range} \\  \text{classical} \phantom{ab} }} = \dfrac{4 \pi G}{|\boldsymbol{q}|^2} \dfrac{m_1^2 m_2^2 c^4 - 2(p_1 {\cdot} p_3)^2}{E_1 E_3} \dfrac{1}{ \frac{(\boldsymbol{p} {\cdot} \hat{\boldsymbol{q}})^2}{E_1 E_3} + 1 } + \frac{A \big(\boldsymbol{p}^2, (\boldsymbol{p} {\cdot} \boldsymbol{q})\big)}{|\boldsymbol{q}|}~,
	\end{split}
\end{equation}
where $\hat{\boldsymbol{q}}=\frac{\boldsymbol{q}}{q}$.
In the second term on the right hand side of \eqref{StandardFrameAmp} $A$ is a function of $\boldsymbol{p}^2$ and $(\boldsymbol{p} {\cdot} \boldsymbol{q})$ is explicitly given in Appendix \ref{AppA}. By checking its expression in the NR limit one discovers a series of terms proportional to odd powers of $(\boldsymbol{p} {\cdot} \boldsymbol{q})$. These, together with the $|\boldsymbol{q}|$ in the denominator, lead to ratios that once Fourier transformed result in imaginary terms.
For example at order $O(1/c^2)$, a contribution of the following type occurs
\begin{equation}
\int \frac{d^3 \boldsymbol{q}}{(2 \pi )^3} \, \frac{\boldsymbol{p} \cdot \boldsymbol{q}}{\boldsymbol{q}^2} \, e^{-i \boldsymbol{q} \cdot \boldsymbol{r}} = i \frac{\boldsymbol{p} \cdot \boldsymbol{r}}{4 \pi r^3}~.
\end{equation}
The resulting imaginary terms are not surprising in a scattering amplitude, but they cannot be accepted as parts of a classical potential. We then have to look for a choice of the momenta alternative to (\ref{Standardframe}) that could solve this issue. To this end we repeat the computation above in a generically parameterised reference frame:
\begin{equation}
	\begin{cases} 
	\boldsymbol{p}_1 = - \boldsymbol{p}_3 = \boldsymbol{p} +  a \boldsymbol{q} \\ 
	\boldsymbol{p}_2 = - \boldsymbol{p}_4 = \boldsymbol{p} + (a - 1) \boldsymbol{q}
	\end{cases},
\end{equation}
where $a$ is a free real parameter. In doing so, we retrieve the amplitude (\ref{StandardFrameAmp}) but with a difference, an overall factor $(2a-1)$ in front of the term proportional to $A$. Clearly this means that we can get rid of all the imaginary terms mentioned above by simply setting $a=1/2$. Therefore the final expression for the 3-momenta is
\begin{equation}
\begin{cases} 
\boldsymbol{p}_1 = - \boldsymbol{p}_3 = \boldsymbol{p} + \boldsymbol{q}/2 \\ 
\boldsymbol{p}_2 = - \boldsymbol{p}_4 = \boldsymbol{p} - \boldsymbol{q}/2
\end{cases},
\end{equation}
with the $q$-dependence now equally distributed to initial and final momenta.
Accordingly, the new amplitude reads
\begin{equation}
	\bigg[M^{\text{tree}}(\boldsymbol{q})\bigg]_{\substack{
			\text{long-range} \\  \text{classical} \phantom{ab} }} = \dfrac{4 \pi G}{|\boldsymbol{q}|^2} \bigg[ \dfrac{m_1^2 m_2^2 c^4 - 2(p_1 {\cdot} p_3)^2}{E_1 E_3} \dfrac{1}{ \frac{(\boldsymbol{p} {\cdot} \hat{\boldsymbol{q}})^2}{E_1 E_3} + 1 }\bigg]_{|\boldsymbol{q}|=0}.
\end{equation}
This is precisely the first term of (\ref{StandardFrameAmp}), where now the quantities inside the square brackets are evaluated at $|\boldsymbol{q}|=0$ since in the new reference frame, unlike in the previous one, 4-momenta and energies actually depend on $\boldsymbol{q}$.

As usual the next step is to Fourier transform.
Aside from all the $\boldsymbol{q}$-independent quantities that can be moved outside the integral, this means to evaluate
\begin{equation} \label{FintCM}
\int \frac{d^3 \boldsymbol{q}}{(2 \pi )^3} \, \frac{1}{|\boldsymbol{q}|^2} \frac{e^{-i \boldsymbol{q} \cdot \boldsymbol{r}}}{1 + \alpha (\boldsymbol{p} \cdot \hat{\boldsymbol{q}})^2}~,
\end{equation}
where for the sake of compactness we have defined:
\begin{equation}
 	\alpha \equiv \frac{c^2}{[E(p_1) E(p_3)]_{|\boldsymbol{q}|=0}} = \frac{1}{\sqrt{\boldsymbol{p}^2 + m_1^2 c^2} \sqrt{\boldsymbol{p}^2 + m_2^2 c^2}}~.
\end{equation}
 
Then, by noting that
\begin{equation} \label{minor}
	\alpha (\boldsymbol{p} \cdot \hat{\boldsymbol{q}})^2 = \frac{(\boldsymbol{p} \cdot \hat{\boldsymbol{q}})^2}{\sqrt{\boldsymbol{p}^2 + m_1^2 c^2} \sqrt{\boldsymbol{p}^2 + m_2^2 c^2}} \leq \frac{|\boldsymbol{p}|^2}{\sqrt{\boldsymbol{p}^2 + m_1^2 c^2} \sqrt{\boldsymbol{p}^2 + m_2^2 c^2}} < 1
\end{equation}
we can rewrite the integrand in (\ref{FintCM}) using
\begin{equation}
	\frac{1}{1+ \alpha(\boldsymbol{p} \cdot \hat{\boldsymbol{q}})^2} = \sum_{n=0}^{+ \infty} (-1)^n \alpha^n (\boldsymbol{p} \cdot \hat{\boldsymbol{q}})^{2n}
\end{equation}
and evaluate the Fourier transform term by term:
\begin{equation} \label{FTbellina}
	\bigg[\frac{(\boldsymbol{p} \cdot \hat{\boldsymbol{q}})^{2n}}{|\boldsymbol{q}|^2}\bigg]_{\text{FT}} \equiv \int \frac{d^3 \boldsymbol{q}}{(2 \pi )^3} \, \frac{(\boldsymbol{p} \cdot \hat{\boldsymbol{q}})^{2n}}{|\boldsymbol{q}|^2} e^{-i \boldsymbol{q} \cdot \boldsymbol{r}} = \frac{1}{4 \pi r } \frac{(2n)!}{2^{2n} (n!)^2}\big[\boldsymbol{p}^2-(\boldsymbol{p} \cdot \hat{\boldsymbol{r}})^2 \big]^n~,
\end{equation}
for a non-negative integer $n$.
In this way we find
\begin{equation}
\setlength{\jot}{10pt}
	\begin{split}
	\int \frac{d^3 \boldsymbol{q}}{(2 \pi )^3} \, \frac{1}{|\boldsymbol{q}|^2} \frac{e^{-i \boldsymbol{q} \cdot \boldsymbol{r}}}{1 + \alpha (\boldsymbol{p} \cdot \hat{\boldsymbol{q}})^2} & = \frac{1}{4 \pi r} \sum_{n=0}^{+\infty}  (-1)^n  \frac{(2n)!}{(n!)^2}\bigg[\alpha \frac{\boldsymbol{p}^2-(\boldsymbol{p} \cdot \hat{\boldsymbol{r}})^2}{4}\bigg]^n = \\
	& = \frac{1}{4 \pi r} \frac{1}{\sqrt{1+ \alpha [\boldsymbol{p}^2 - (\boldsymbol{p}\cdot \hat{\boldsymbol{r}})^2 \big]}}~.
	\end{split}
\end{equation}
The last equality holds since
\begin{equation}
	\sum_{n=0}^{+\infty}  (-1)^n \frac{(2n)!}{(n!)^2} z^n = \frac{1}{\sqrt{1+4z}} \, \, \, \, \text{for} \, \, |z|<\frac{1}{4}
\end{equation}
and, just like it was shown in (\ref{minor}), we have
\begin{equation}
	\alpha \big[ \boldsymbol{p}^2 - (\boldsymbol{p}\cdot \hat{\boldsymbol{r}}) \big] < 1~.
\end{equation}

In conclusion our final result for the position space 1PM potential in the center of mass frame is
\begin{equation} \label{Result}
	V_{1\text{PM}}(\boldsymbol{p}, \boldsymbol{r}) = - \frac{G}{r} \bigg[\frac{2(p_1 \cdot p_3)^2 - m_1^2 m_2^2 c^4}{E(p_1) E(p_3)} \frac{1}{\sqrt{1 + \frac{c^2}{E(p_1) E(p_3)} \big[ \boldsymbol{p}^2 - (\boldsymbol{p}\cdot \hat{\boldsymbol{r}})^2 \big]} }\bigg]_{|\boldsymbol{q}|=0}~.
\end{equation}
Interestingly, we can rewrite \eqref{Result} as
\begin{equation}
	V^{\text{new}}_{1\text{PM}} = V^{\text{lit.}}_{1\text{PM}} \times \left(1+\dfrac{\boldsymbol{p}^2-(\boldsymbol{p} \cdot \hat{\boldsymbol{r}})^2}{\sqrt{\boldsymbol{p}^2+m_1^2 c^2}\sqrt{\boldsymbol{p}^2+m_2^2 c^2}} \right)^{-\frac{1}{2}}
\end{equation}
where $V^{\text{lit.}}_{1\text{PM}} $ is the known 1PM potential (\ref{Vlett}).
Moreover, thanks to the dimensionless additional factor multiplying $V^{\text{lit.}}_{1\text{PM}}$, the NR expansion now gives 
\begin{equation}
	V_{1\text{PM}}(\boldsymbol{p}, \boldsymbol{r}) = - \frac{G m_1 m_2}{r} - \frac{G m_1 m_2}{2rc^2} \bigg[\frac{3 \boldsymbol{p}^2}{m^2_1} + \frac{3 \boldsymbol{p}^2}{m^2_2} + \frac{7\boldsymbol{p}^2}{ m_1 m_2} + \frac{(\boldsymbol{p} \cdot \hat{\boldsymbol{r}})^2}{m_1 m_2} \bigg] + O(1/c^4)
\end{equation}
which exactly reproduces the $O(G)$ terms of the 1PN Einstein-Infeld-Hoffmann potential (\ref{EIHpot}). This is enough to reestablish the complete matching with the 1PN potential as a whole, since the static term
\begin{equation*}
	\frac{G^2m_1 m_2(m_1 + m_2)}{2r^2 c^2}
\end{equation*}
is already correctly reproduced by every 2PM potential derived with the scattering amplitude method in the literature~\cite{Cheung:2018wkq,Bern:2019crd,Cristofoli:2019neg}. The new 1PM potential (\ref{Result}) represents the main result of this Letter.

\section{NR limit beyond the 1PN order}

The extension of this convenient direct match between the NR limit of the 1PM potential (\ref{Result}) and the PN results available in the literature to include higher order in the PN expansion is not however immediately obvious and requires further investigation.

From ref.~\cite{Damour:1988mr}, the paper which first presented a 2PN two-body Hamiltonian, we can extract the $O(G/c^4)$ component of the 2PN potential in the center of mass frame which is given by
\begin{equation} \label{2PNpot}
	V_{2\text{PN}}^{[O(G/c^4)]} = \frac{G m_1 m_2}{8 r c^4} \bigg(\frac{5 \boldsymbol{p}^4}{m_1^4}-\frac{13 \boldsymbol{p}^4}{m_1^2 m_2^2}-\frac{3 (\boldsymbol{p} {\cdot} \hat{\boldsymbol{r}} )^4}{m_1^2 m_2^2}-\frac{2 \boldsymbol{p}^2 (\boldsymbol{p} {\cdot} \hat{\boldsymbol{r}} )^2}{m_1^2 m_2^2}+\frac{5 \boldsymbol{p}^4}{m_2^4} \bigg)~.
\end{equation}
By comparing the non-relativistic expansion of our 1PM potential with (\ref{2PNpot}), the following discrepancy arises:
\begin{equation} \label{D4}
	\begin{split}
	D_4 \equiv V_{1\text{PM}}^{[O(G/c^4)]} - V^{[O(G/c^4)]}_{2\text{PN}} & = \frac{G}{2 c^4 r} \bigg(\frac{1}{m_1}+ \frac{1}{m_2}\bigg)^2 \big[ \boldsymbol{p}^4 - \boldsymbol{p}^2 (\boldsymbol{p} \cdot \hat{\boldsymbol{r}})^2\big] = \\
	& = \frac{4 \pi G}{ c^4} \bigg(\frac{1}{m_1}+ \frac{1}{m_2}\bigg)^2 \bigg[\frac{(\boldsymbol{p} \cdot \hat{\boldsymbol{q}})^2}{|\boldsymbol{q}|^2}\bigg]_{\text{FT}} \boldsymbol{p}^2.
	\end{split}
\end{equation}

We haven't yet been able to find any obvious adjustment of our 1PM potential such that the discrepancy (\ref{D4}) end up being resolved. Very likely, since PN results are not found in the harmonic gauge, here the problem lies in the gauge fixing choice. We discuss this issue further in the \nameref{Conc}.

What we could do for the time being (for now) is to check) whether and how the correct NR limit of our 1PM potential could be obtained by performing some kind of {\it a posteriori} canonical transformation. We have done this up to the 4PN order. Clearly such a transformation should not be restricted to the potential only, but extended also to the other component of our effective Hamiltonian, the free-particle one. The latter is precisely the part of the Hamiltonian which could in principle provide the desired correction. Indeed, since our Hamiltonian reproduces the expected NR limit up to the 1PN order, the transformation we look for has to be proportional, at least, to $G/c^4$:
\begin{itemize}
	\item if a factor $G$ is not included, the NR limit of the free-particle energy results is unavoidably spoiled.
	\item if at least a factor $1/c^4$ is not included, the Newtonian and the 1PN parts of the potential are spoiled.
\end{itemize}
Therefore the additional $O(G)$ terms that we need in order to fix the NR behaviour of our potential could only come from the canonically transformed $O(G^0)$ part of our Hamiltonian. Bearing this in mind, the canonical transformation which resolves the discrepancy (\ref{D4}) relative to the 2PN order is uniquely determined:
\begin{equation} \label{canon}
	\begin{cases}
	\boldsymbol{p} \rightarrow \big[ 1 - A_4(\boldsymbol{p}, \boldsymbol{r}) \big] \boldsymbol{p} \\
	\boldsymbol{r} \rightarrow \big[ 1 + A_4(\boldsymbol{p}, \boldsymbol{r}) \big] \boldsymbol{r}
	\end{cases}
\end{equation} 
where the scalar function $A_4(\boldsymbol{p}, \boldsymbol{r})$ in terms of the Fourier transforms (\ref{FTbellina}) reads
\begin{equation} \label{A4}
	A_4(\boldsymbol{p}, \boldsymbol{r})=\frac{4 \pi G}{ c^4} \bigg(\frac{1}{m_1}+ \frac{1}{m_2}\bigg)^2 \bigg[\frac{(\boldsymbol{p} \cdot \hat{\boldsymbol{q}})^2}{|\boldsymbol{q}|^2}\bigg]_{\text{FT}}
\end{equation}

The matching with the 3PN and 4PN potentials found, for example, in ref.s~\cite{Jaranowski:2015lha,Damour:2014jta} can be restored with the exact same procedure: starting from the $n$PN discrepancy
\begin{equation}
	D_{2n}  \equiv V_{1\text{PM}}^{[O(G/c^{2n})]} - V^{[O(G/c^{2n})]}_{n\text{PN}}
\end{equation}
we find the respective scalar function $A_{2n}(\boldsymbol{p}, \boldsymbol{r})$ that should be added to the canonical transformation (\ref{canon}). Our final result is
\begin{equation} \label{FinalTRansf}
	\begin{cases}
	\displaystyle \boldsymbol{p} \rightarrow \big[ 1 -  \sum_{n=2}^{4} A_{2n}(\boldsymbol{p}, \boldsymbol{r}) \big] \boldsymbol{p} \\
	\displaystyle \boldsymbol{r} \rightarrow \big[ 1 + \sum_{n=2}^{4} A_{2n}(\boldsymbol{p}, \boldsymbol{r}) \big] \boldsymbol{r}
	\end{cases}
\end{equation}
where, in addition to (\ref{A4}), we find
\begin{equation}
	A_6(\boldsymbol{p}, \boldsymbol{r}) = -\frac{2 \pi G}{c^6} \bigg(\frac{1}{m_1^3}+\frac{1}{m_2^3}\bigg) \bigg[\frac{(\boldsymbol{p} \cdot \hat{\boldsymbol{q}})^2}{|\boldsymbol{q}|^2}\bigg]_{\text{FT}} \boldsymbol{p}^2 - \frac{4 \pi G}{c^6} \frac{1}{m_1 m_2} \bigg(\frac{1}{m_1}+\frac{1}{m_2}\bigg)\bigg[\frac{(\boldsymbol{p} \cdot \hat{\boldsymbol{q}})^4}{|\boldsymbol{q}|^2}\bigg]_{\text{FT}}
\end{equation}
and
\begin{equation}
\setlength{\jot}{10pt}
	\begin{split} 
		A_8(\boldsymbol{p}, \boldsymbol{r}) & = - \frac{\pi G }{c^8} \frac{1}{m_1 m_2 (m_1 + m_2)} \bigg(\frac{1}{m_1^2}+\frac{1}{m_2^2}\bigg)\bigg\{ \frac{7}{8} \bigg[\frac{1}{|\boldsymbol{q}|^2}\bigg]_{\text{FT}} \boldsymbol{p}^6 + \frac{1}{4} \bigg[\frac{(\boldsymbol{p} \cdot \hat{\boldsymbol{q}})^2}{|\boldsymbol{q}|^2}\bigg]_{\text{FT}} \boldsymbol{p}^4 \bigg\} + \\
		& + \frac{\pi G}{c^8} \frac{1}{m_1^2 m_2^2} \bigg(\frac{1}{m_1} + \frac{1}{m_2}\bigg) \bigg\{4 \bigg[\frac{(\boldsymbol{p} \cdot \hat{\boldsymbol{q}})^4}{|\boldsymbol{q}|^2}\bigg]_{\text{FT}} \boldsymbol{p}^2 + 6 \bigg[\frac{(\boldsymbol{p} \cdot \hat{\boldsymbol{q}})^6}{|\boldsymbol{q}|^2}\bigg]_{\text{FT}} - 2 \bigg[\frac{(\boldsymbol{p} \cdot \hat{\boldsymbol{q}})^8}{|\boldsymbol{q}|^2}\bigg]_{\text{FT}} \frac{1}{\boldsymbol{p}^2} \bigg\} + \\
		& + \frac{3 \pi G}{2 c^8} \bigg(\frac{1}{m_1^5}+\frac{1}{m_2^5}\bigg) \bigg[\frac{(\boldsymbol{p} \cdot \hat{\boldsymbol{q}})^2}{|\boldsymbol{q}|^2}\bigg]_{\text{FT}} \boldsymbol{p}^4  - \frac{4 \pi G}{c^8} \frac{1}{m_1^2 m_2^2 (m_1 + m_2)}\bigg[\frac{(\boldsymbol{p} \cdot \hat{\boldsymbol{q}})^2}{|\boldsymbol{q}|^2}\bigg]_{\text{FT}} \boldsymbol{p}^4
	\end{split}
\end{equation}

We studied the transformation (\ref{FinalTRansf}) also under two particular choices for the masses $m_1$ and $m_2$. This is the content of Appendix \ref{AppB}.

\section{1PM Scattering Angle With the New Potential}

Starting from an effective Hamiltonian one can always obtain physical observables. Indeed they must remain unaffected under any non-physical modification, thus providing the perfect context to test the physical consistency of our newly found expression for the 1PM potential given in eq.~\eqref{Result}, compared to the one normally used in the literature. Along these lines, in this section we will compute the 1PM scattering angle from our potential \eqref{Result} and show that the resulting expression is consistent with the known scattering angle that can be found in the literature~\cite{Sugar:1969rn,Paliov:1974fu,Wallace:1973iu,KoemansCollado:2019ggb,Bjerrum-Bohr:2019kec,Kalin:2020mvi}.

\subsection{The Scattering Angle}

Since in our setup the spin is not involved, we can restrict ourselves to the case where the motion is lying on a plane. Accordingly, the momentum can be decomposed as
\begin{equation} \label{Decomposition}
	\boldsymbol{p}^2 = \boldsymbol{p}_r^2 + \frac{\boldsymbol{L}^2}{r^2}~,
\end{equation}
in terms of the radial momentum $\boldsymbol{p}_r$ and the conserved angular momentum $\boldsymbol{L}$. Once $\boldsymbol{p}_r$ is known, the scattering angle follows from the formula~\cite{Damour:2016gwp,Cristofoli:2019neg}:
\begin{equation} \label{SCA}
	\chi = -\pi -2 \int_{r_{\text{min}}}^{+ \infty} dr \, \frac{\partial p_r}{\partial L}~,
\end{equation}
where $r_{\text{min}}$ corresponds to the minimum distance between the two scattering bodies.
Denoting with $E$ the total conserved energy, we can extract the expression for the radial momentum at order $O(G)$ by solving perturbatively the equation
\begin{equation}
	H(\boldsymbol{r}, \boldsymbol{p}) = c \sqrt{\boldsymbol{p}^2 + m_1^2} + c \sqrt{\boldsymbol{p}^2 + m_2^2} + V_{1\text{PM}}(\boldsymbol{r}, \boldsymbol{p}^2) = E   \qquad (\text{Energy Conservation})
\end{equation}
together with the momentum decomposition (\ref{Decomposition}).
In this way we find
\begin{equation} \label{pr}
	p_r^2(r,L) = p^2_{\infty} - \frac{L^2}{r^2} + P^{(1)}\frac{G}{r} + O(G^2)~,
\end{equation}
where we defined the momentum at infinity, $p_{\infty}$, such that
\begin{equation}
	E = c \sqrt{p_{\infty}^2 + m_1^2} + c \sqrt{p_{\infty}^2 + m_2^2}~,
\end{equation}
and the 1PM coefficient can be derived from our expression for the 1PM potential \eqref{Result} and is given by
\begin{eqnarray}\label{P1}
	P^{(1)} \equiv &&\frac{4 \big(p^2_{\infty}+ \sqrt{p_{\infty}^2 + m_1^2 c^2} \sqrt{p_{\infty}^2 + m_2^2 c^2} \big)^2 - 2 m_2^2 m_2^2 c^4 }{E^2} \cr
	&&\times \bigg(1+ \dfrac{L^2}{\sqrt{p_{\infty}^2 + m_1^2 c^2} \sqrt{p_{\infty}^2 + m_2^2 c^2} \, r^2}\bigg)^{-\frac{1}{2}}~.
\end{eqnarray}
We note that at this stage the new potential \eqref{Result} modifies only the expression of $P^{(1)}$ which differs from the one used in the literature just for the multiplicative adimensional factor we isolated in the second line of (\ref{P1}). Nevertheless, this difference has a remarkable impact in the computation of the scattering angle \eqref{SCA} because the new factor features both an $L$ and an $r$ dependence. The first one is relevant for the integrand in (\ref{SCA}),
\begin{equation}
	\frac{\partial p_r}{\partial L} = - \frac{L^2}{r^2 p_r(r)} + \frac{G}{2 r p_r(r)}\frac{\partial P^{(1)}}{\partial L}~,
\end{equation}
while the $r$ dependence affects the position of the inversion point $r_{\text{min}}$. In fact the condition
\begin{equation}
	p_r(r_{\text{min}}) = 0 
\end{equation}
from which  $r_{\text{min}}$ is usually derived as the positive root, has now 6 solutions, and only a perturbative treatment allows us to identify it. At order $G$ we find
\begin{equation}
	r_{\text{min}} = r_{0} - G \frac{r_0 P^{(1)}_{\infty}}{2 p_{\infty}^2 \sqrt{r_0^2+(\alpha_{\infty}L)^2}}~,
\end{equation}
compactly written in terms of
\begin{equation}
\begin{split}
r_{0} & \equiv \frac{L}{p_{\infty}}~, \quad P^{(1)}_{\infty} \equiv \frac{4 \big(p^2_{\infty}+ \sqrt{p_{\infty}^2 + m_1^2 c^2} \sqrt{p_{\infty}^2 + m_2^2 c^2} \big)^2 - 2 m_2^2 m_2^2 c^4 }{E^2}~, \\
\alpha_{\infty} & \equiv \frac{1}{\sqrt{p_{\infty}^2 + m_1^2 c^2} \sqrt{p_{\infty}^2 + m_2^2 c^2 } }~.
\end{split}	
\end{equation}
Using this notation and the expression of $p_r$ given in (\ref{pr}) we get to the following result for the scattering angle
\begin{equation}
	\chi = - \pi + 2 r_0 I_1 + G r_0 P^{(1)}_{\infty} \alpha_{\infty} I_2
\end{equation}
where the integrals
\begin{equation}
I_1 \equiv \int_{r_{\text{min}}}^{+ \infty} dr \, \frac{1}{r \sqrt{r^2 - r_0^2 + G \dfrac{P^{(1)}_{\infty} r^2}{p_{\infty}^2 \sqrt{r^2 + (\alpha_{\infty}L)^2}}}}
\end{equation}
and
\begin{equation}
I_2 \equiv \int_{r_{\text{min}}}^{+ \infty} dr \, \frac{r}{\bigg(r^2 + (\alpha_{\infty}L)^2 \bigg)^\frac{3}{2} \sqrt{r^2 - r_0^2 + G \dfrac{P^{(1)}_{\infty} r^2}{ p_{\infty}^2 \sqrt{r^2 + (\alpha_{\infty}L)^2}}}}~,
\end{equation}
have to be evaluated respectively at order $O(G)$ and $O(G^0)$. 

In both these computations we used
\begin{equation}
	\int_{g(G)}^{+ \infty} dr \, f(r, G) = \int_{g(0)}^{+ \infty} dr \, f(r, 0) + G \left(  \frac{\partial}{\partial G} \int_{g(G)}^{+ \infty} dr \, f(r, G) \right) \Bigg\rvert_{G=0} + O(G^2)
\end{equation}
and
\begin{equation}
	\frac{\partial}{\partial G}  \int_{g(G)}^{+ \infty} dr \, f(r, G) = - f\big(g(G), G \big) \frac{\partial}{\partial G} g(G) + \int_{g(G)}^{+ \infty} dr \, \frac{\partial}{\partial G} f(r, G)~.
\end{equation}
Carefully handling some divergences that cancel out among the various terms of these integrals, we find
\begin{equation}
	\begin{split}
	I_1 & = \frac{\pi}{2 r_0} + G\frac{P^{(1)}_{\infty}}{2 p_{\infty}^2 \big(r_0^2 + (\alpha_{\infty}L)^2 \big)} + O(G^2)~,\\
	I_2 & = \frac{1}{r_0^2 + (\alpha_{\infty}L)^2}+ O(G)~.
	\end{split}
\end{equation}
Therefore our final result for the scattering angle is
\begin{equation}
	\chi = \frac{G P^{(1)}_{\infty} r_0 }{p_\infty^2 \big(r_0^2 + (\alpha_{\infty}L)^2 \big)} + \frac{G P^{(1)}_{\infty} r_0 \alpha_{\infty}^2}{r_0^2 + (\alpha_{\infty}L)^2} + O(G^2) = \frac{G P^{(1)}_{\infty}}{L p_\infty} + O(G^2)
\end{equation}
which is in perfect agreement with the one computed in refs.\ \cite{Sugar:1969rn,Paliov:1974fu,Wallace:1973iu,KoemansCollado:2019ggb,Bjerrum-Bohr:2019kec,Kalin:2020mvi}.

\section{Conclusion}
\label{Conc}

In this paper we showed how the reference frame dependence in the derivation of the 1PM potential enables one to find a new expression for the 1PM potential that generalizes what was previously found in the literature. This results in an additional dimensionless factor that should be multiplied to the standard 1PM potential and that, in the non relativistic limit, directly and correctly reproduces the Einstein-Infeld-Hoffman 1PN potential.

This nice feature is not immediately enjoyed by the higher PN orders and, at this stage, we have been able to provide a canonical transformation from which one reproduces the PN potential up to the 4PN order, starting with the 1PM potential \eqref{Result}. A future goal would be to provide an expression for the PM potential which, once expanded for small velocities, gives automatically the known result in the PN expansion, without the need of performing a canonical transformation.
This would indeed be an interesting task, although highly non-trivial: the PN results~\cite{Damour:1988mr,Jaranowski:2015lha,Damour:2014jta} we used as a reference are all obtained in ADM coordinates. Moreover in ref.~\cite{Damour:1988mr} it was shown that the approach that uses harmonic coordinates leads to a generalized acceleration-dependent Lagrangian, which prevents the derivation of a proper associated Hamiltonian. Therefore we believe that a possible resolution of this issue can be traced back to the gauge choice. In this respect it should be mentioned that the possibility of directly reproducing PN results beyond the first order is likely lost already when one employs the linear expanded metric (\ref{LinerizedMetric}). That is because such a choice limits the gauge fixing freedom from the start to just the linearized gauges, which seem unsuitable for the purpose at hand. All this considered, a resolution of this issue could require the development of an alternative quantization procedure instead of the customary one we have adopted here.

Nevertheless, by reproducing the correct 1PM scattering angle we have performed a consistency check of our extended result for the 1PM potential that shows its physical equivalence to the 1PM potential found previously in the literature~\cite{Bern:2019crd,Cristofoli:2019neg,PhysRevD.100.066028}.

\begin{acknowledgments}

We thank S. Mezzasoma  and F. Camilloni for their contribution in an early stage of this project. We thank N. E. Bjerrum-Bohr for reading the manuscript and for useful comments. 
T.~H.~acknowledges support from the Independent Research Fund Denmark grant number DFF-6108-00340 ``Towards a deeper understanding of black holes with non-relativistic holography". G.~G. and M.~O.~acknowledge support from the project ``Black holes, neutron stars and gravitational waves" financed by Fondo Ricerca di Base 2018 of the University of Perugia. T.~H.~thanks Perugia University and G.~G. and M.~O.~thank Niels Bohr Institute for hospitality.
\end{acknowledgments}

\appendix

\section{The function $A$}
\label{AppA}

In this appendix we give the explicit expression for the function $A \big(\boldsymbol{p}^2, (\boldsymbol{p} {\cdot} \boldsymbol{q})\big)$ that appears in Eq.\ (\ref{StandardFrameAmp}). We can write
\begin{equation}
	A \big(\boldsymbol{p}^2, (\boldsymbol{p} {\cdot} \boldsymbol{q})\big) = \dfrac{\text{Num}A}{\text{Den}A}
\end{equation}
where
\begin{equation}
\begin{split}
& \text{Num}A = 2 \pi G (\boldsymbol{p} {\cdot} \boldsymbol{q}) \bigg( 3 m_1^2 m_2^2 c^4 + 2 \boldsymbol{p}^2 (m_1^2 + m_2^2) c^2 \bigg) \bigg\{ \bigg[ 2 m_1^2 m_2^2 c^4 + (m_1^2 + m_2^2)E_1 E_3 + \\
& + 2 \boldsymbol{p}^2\bigg(\frac{E_1 E_3}{c^2} + (m_1^2 + m_2^2)c^2\bigg) + 2 \boldsymbol{p}^4\bigg] + 4 (\boldsymbol{p} {\cdot} \boldsymbol{q})^2 \bigg[ \boldsymbol{p}^4\bigg( 4 \frac{E_1 E_3}{c^2} + 5(m_1^2 + m_2^2)c^2 \bigg) + \\
& + m_1^2 m_2^2 c^4 \bigg(2 \frac{E_1 E_3}{c^2} + (m_1^2 + m_2^2)c^2 \bigg) + \boldsymbol{p}^2 \bigg( (m_1^4 + 6 m_1^2 m_2^2 + m_2^4) c^4 + 3(m_1^2 + m_2^2) E_1 E_3 \bigg) + \\ 
& + 4 \boldsymbol{p}^6\bigg] \bigg\}~,
\end{split}
\end{equation}

\begin{equation}
	\text{Den}A = \frac{(E_1 E_3)^2}{c^2} \bigg(\frac{E_1 E_3}{c^2} + (\boldsymbol{p} {\cdot} \boldsymbol{q})^2 \bigg)^2~.
\end{equation}

\section{The canonical transform under particular conditions}
\label{AppB}
In this appendix we will present how the canonical transformation \eqref{FinalTRansf} simplifies under two specific choices for the masses $m_1$ and $m_2$ of the astrophysical bodies in the binary system.

\subsection{Case $m_1 \gg m_2$}

Here we consider the case where one of the two bodies has a mass much bigger than the other. Because of the symmetry of our Hamiltonian under the exchange of the two objects, that it is clearly reflected also in the transformation (\ref{FinalTRansf}), it is not important which one of the two masses is assumed to be the biggest. We arbitrarily choose it to be $m_1$. The scalar functions $A_{2n}(\boldsymbol{p}, \boldsymbol{r})$ become

\begin{equation}
A_{4}(\boldsymbol{p}, \boldsymbol{r})= \frac{4 \pi G}{c^4} \frac{1}{m_2} \bigg[\frac{(\boldsymbol{p} \cdot \hat{\boldsymbol{q}})^2}{|\boldsymbol{q}|^2}\bigg]_{\text{FT}},
\end{equation}

\begin{equation}
A_{6}(\boldsymbol{p}, \boldsymbol{r})= - \frac{2 \pi G}{c^6} \frac{1}{m_2^3} \bigg[\frac{(\boldsymbol{p} \cdot \hat{\boldsymbol{q}})^2}{|\boldsymbol{q}|^2}\bigg]_{\text{FT}}\boldsymbol{p}^2 - \frac{4 \pi G}{c^6} \frac{1}{m_1 m_2^2} \bigg[\frac{(\boldsymbol{p} \cdot \hat{\boldsymbol{q}})^4}{|\boldsymbol{q}|^2}\bigg]_{\text{FT}},
\end{equation}

\begin{equation}
\setlength{\jot}{10pt}
\begin{split}
A_{8}(\boldsymbol{p}, \boldsymbol{r}) & = - \frac{\pi G}{c^8} \frac{1}{m_1^2 m_2^3} \bigg\{ \frac{7}{8} \bigg[\frac{1}{|\boldsymbol{q}|^2}\bigg]_{\text{FT}} \boldsymbol{p}^6 + \frac{1}{4} \bigg[\frac{(\boldsymbol{p} \cdot \hat{\boldsymbol{q}})^2}{|\boldsymbol{q}|^2}\bigg]_{\text{FT}} \boldsymbol{p}^4 - 4 \bigg[\frac{(\boldsymbol{p} \cdot \hat{\boldsymbol{q}})^4}{|\boldsymbol{q}|^2}\bigg]_{\text{FT}} \boldsymbol{p}^2 - 6 \bigg[\frac{(\boldsymbol{p} \cdot \hat{\boldsymbol{q}})^6}{|\boldsymbol{q}|^2}\bigg]_{\text{FT}} + \\
& + 2 \bigg[\frac{(\boldsymbol{p} \cdot \hat{\boldsymbol{q}})^8}{|\boldsymbol{q}|^2}\bigg]_{\text{FT}} \frac{1}{\boldsymbol{p}^2} \bigg\} + \frac{3 \pi G}{2 c^8} \frac{1}{m_2^5} \bigg[\frac{(\boldsymbol{p} \cdot \hat{\boldsymbol{q}})^2}{|\boldsymbol{q}|^2}\bigg]_{\text{FT}} \boldsymbol{p}^4 - \frac{4 \pi G}{c^8} \frac{1}{m_1^3 m_2^2} \bigg[\frac{(\boldsymbol{p} \cdot \hat{\boldsymbol{q}})^2}{|\boldsymbol{q}|^2}\bigg]_{\text{FT}} \boldsymbol{p}^4.
\end{split}
\end{equation}
Given the imposed constraint on the masses, the fact that the former symmetry $(1\rightleftarrows2)$ is lost should not be a source of concern.

\subsection{Case $m_1 = m_2$}

Finally, we analyse the special case in which the two bodies share the same mass, hence $m_1=m_2 \equiv m$. The scalar functions $A_{2n}(\boldsymbol{p}, \boldsymbol{r})$ greatly simplify and we find
\begin{equation}
A_{4}(\boldsymbol{p}, \boldsymbol{r})= \frac{8 \pi G}{c^4} \frac{1}{m} \bigg[\frac{(\boldsymbol{p} \cdot \hat{\boldsymbol{q}})^2}{|\boldsymbol{q}|^2}\bigg]_{\text{FT}},
\end{equation}

\begin{equation}
A_{6}(\boldsymbol{p}, \boldsymbol{r})= - \frac{8 \pi G}{c^6}  \frac{1}{m^3}\bigg\{ \frac{1}{2}\bigg[\frac{(\boldsymbol{p} \cdot \hat{\boldsymbol{q}})^2}{|\boldsymbol{q}|^2}\bigg]_{\text{FT}}\boldsymbol{p}^2 + \bigg[\frac{(\boldsymbol{p} \cdot \hat{\boldsymbol{q}})^4}{|\boldsymbol{q}|^2}\bigg]_{\text{FT}} \bigg\},
\end{equation}

\begin{equation}
\setlength{\jot}{10pt}
\begin{split}
A_{8}(\boldsymbol{p}, \boldsymbol{r}) & = - \frac{\pi G}{c^8} \frac{1}{m^5} \bigg\{ \frac{7}{8} \bigg[\frac{1}{|\boldsymbol{q}|^2}\bigg]_{\text{FT}} \boldsymbol{p}^6 - \frac{3}{4} \bigg[\frac{(\boldsymbol{p} \cdot \hat{\boldsymbol{q}})^2}{|\boldsymbol{q}|^2}\bigg]_{\text{FT}} \boldsymbol{p}^4 - 8 \bigg[\frac{(\boldsymbol{p} \cdot \hat{\boldsymbol{q}})^4}{|\boldsymbol{q}|^2}\bigg]_{\text{FT}} \boldsymbol{p}^2 + \\
& - 12 \bigg[\frac{(\boldsymbol{p} \cdot \hat{\boldsymbol{q}})^6}{|\boldsymbol{q}|^2}\bigg]_{\text{FT}} + 4 \bigg[\frac{(\boldsymbol{p} \cdot \hat{\boldsymbol{q}})^8}{|\boldsymbol{q}|^2}\bigg]_{\text{FT}} \frac{1}{\boldsymbol{p}^2}   \bigg\}.
\end{split}
\end{equation}

\bibliographystyle{ieeetr}
\bibliography{scibib}
\clearpage
\end{document}